\documentclass[twocolumn,showpacs,nofootinbib]{revtex4}
\usepackage{epsf}
\usepackage{dcolumn}

\newcommand{\BE}{\begin{equation}}
\newcommand{\EE}{\end{equation}}
\newcommand{\BA}{\begin{eqnarray}}
\newcommand{\EA}{\end{eqnarray}}
\def\be{\begin{equation}}
\def\ee{\end{equation}}
\def\bea{\begin{eqnarray}}
\def\eea{\end{eqnarray}}

\def\la{\mathrel{\mathpalette\fun <}}

\def\fun#1#2{\lower3.6pt\vbox{\baselineskip0pt\lineskip.9pt
  \ialign{$\mathsurround=0pt#1\hfil##\hfil$\crcr#2\crcr\sim\crcr}}}
\begin{document}
\input epsf
\draft
\renewcommand{\topfraction}{0.8}
\preprint{astro-ph/0312183, \today}
\title{\bf\LARGE Testing the Cosmological Constant\\ as a Candidate  for Dark Energy}

\author{\bf Jan Kratochvil, Andrei Linde}
\affiliation{ { Department
  of Physics, Stanford University, Stanford, CA 94305-4060,
USA}    }
\author{\bf Eric V. Linder}
\affiliation{Physics Division, Lawrence Berkeley National Laboratory,  Berkeley, California 94720, USA}
\author{\bf Marina Shmakova}
\affiliation{ Stanford Linear Accelerator Center, Stanford University,
Stanford CA 94309, USA}

{\begin{abstract} It may be difficult to single out the best model
of dark energy on the basis of the existing and planned
cosmological observations, because the parameter space of many
different models can lead to similar observational consequences.
However, each particular model can be studied and either found
consistent with observations or ruled out. In this paper, we
concentrate on the possibility to test and rule out the simplest
and by far the most popular of the models of dark energy, the
theory described by general relativity with positive vacuum energy
(the cosmological constant). We evaluate the conditions under
which this model could be ruled out by the future observations
made by the Supernova/Acceleration Probe SNAP (both for supernovae
and weak lensing) and by the Planck Surveyor cosmic microwave
background satellite.

\end{abstract}}
\pacs{98.80.Cq, 11.25.-w, 04.65.+e}
\maketitle
\section{\label{Introduction}Introduction}

The observed present acceleration of the expansion of the universe
\cite{supernova,Bond,Spergel:2003cb} is commonly attributed to the
presence of dark energy throughout the universe and gives rise to
the question of what this dark energy actually is. There exist
many different models of dark energy. Some of them are based on a
particular choice of the scalar field potential
\cite{Banks,Linde1986,dark}, whereas some other models are based
on certain modifications of general relativity, see e.g.
\cite{Mukohyama:2003nw} and references therein. Many of these
models can have very similar observational consequences for
certain choices of their parameters. Therefore it would be
extremely hard to determine exactly which model, if any, is
correct. This has led some authors to express a rather pessimistic
attitude towards the observational investigation of dark energy
\cite{Maor:2001ku}.

However, one can look at this issue from a different perspective.
Instead of trying to find which model of dark energy is correct,
one can try to find which models can be ruled out by the existing
and planned observations. This goal is quite realistic, and it
can bring us extremely important information about the
fundamental physics.

One particular case is especially interesting: the theory of dark
energy based on general relativity with a constant vacuum energy
$\Lambda>0$ (positive cosmological constant). First of all, the
$\Lambda$CDM model is by far the simplest dark energy model. In
this model the dark energy remains constant, with the equation of
state, or pressure to density ratio, $w\equiv p/\rho=-1$.
Moreover, at present this is the only known dark energy model
related to M/string theory  \cite{Kachru:2003aw}. Whereas it is
quite possible that M/string theory can describe quintessence with
a time-dependent equation of state, all existing models of this
type have problems describing stable compactifications of internal
dimensions; see a discussion of this issue in
\cite{Kallosh:2002gf,Kachru:2003aw,Gutperle:2003kc}.

An independent argument in favor of the simple cosmological
constant model is obtained if one tries to find an explanation of
the anomalous smallness of the dark energy density $\Lambda$ (the
cosmological constant problem). A possible solution to this
problem can be found in the context of the theory of eternal
inflation if one replaces the cosmological constant by the theory
of dark energy with a flat potential and uses the anthropic principle
\cite{Linde1986,Weinberg:2000yb,Garriga:1999bf,Linde:2002gj}.
However, in the simplest versions of this scenario the slope of
the potential is expected to be so small that all observational
consequences of this theory should be indistinguishable from the
theory with a constant vacuum energy density
\cite{Garriga:2002tq}.

An additional reason to be interested in the possible time
evolution of dark energy is related to inflationary cosmology.
Acceleration of the universe with time-independent vacuum energy
density is similar to the old inflation in the false vacuum
state, whereas acceleration in the universe with the time
dependent vacuum energy is similar to the standard slow-roll
inflation. If we find that the universe experiences a stage of
slow-roll inflation  right now, this would make the slow-roll
inflation in the early universe even more plausible.

On the other hand, if we do not find any observable deviations
from the predictions of the simplest $\Lambda$CDM model, this
will make all models of the dynamically changing dark energy much
less attractive.

Therefore in this paper, instead of studying dark energy models
in general, we will concentrate on a single well defined issue:
What kind of observations could rule out the simple $\Lambda$CDM
model? An answer to this question may help us in planning future
experiments which would be specifically optimized for testing the
simplest cosmological constant model.

The leading edge for dark energy exploration, ever since the
current accelerating expansion of the  universe was discovered
\cite{supernova}, have been Type Ia supernovae observations, and
they will acquire a new state-of-the-art instrument in the form
of the Supernova/Acceleration Probe (SNAP) satellite \cite{snap}.
SNAP will not only perform a distance-redshift measurement of
some 2000 supernovae (SNAP [SN]), but also conduct a wide field weak
gravitational lensing survey (SNAP [WL])
\cite{refregier}, providing a complementary data set, as will at
some point the ground-based LSST survey \cite {lsst}. Additional
measurements of the CMB, provided by the Planck Surveyor cosmic
microwave background satellite \cite {planck}, will help tighten
the constraints obtained by SNAP.

In this paper, we will investigate a possibility to tighten the
constraints on the parameters of dark energy and to test/rule out
the simplest $\Lambda$CDM model using the results of these
experiments. Our approach will be based on the methods developed
in our previous paper \cite{Kallosh:2003bq}.

\

\section{\label{Excluding} Excluding the Cosmological Constant}

The process of excluding the cosmological constant (or any other
model) as a candidate  for dark energy is achieved by mapping
out the expansion history of the universe, i.e.\ the
time-evolution of the scale factor $a(t)$. This is accomplished,
for instance, by measuring the distance-redshift relation of Type
Ia supernovae, which serve as \lq\lq standard candles\rq\rq\ in
cosmology, having a known intrinsic luminosity normalizable
through one parameter based on the width of
their lightcurves, or flux vs.\ time relation. The proposed
dedicated dark energy
satellite mission SNAP, the Supernova/Acceleration Probe, will
measure the precise luminosity distances to approximately 2000
such Type Ia supernovae within 2 years of its launch. The
redshift range of these observed supernovae will span from $z=0.1$
out to $z=1.7$. In our calculations below, we bin the future data
into 17 equally spaced redshift bins, and also include the
expected $\sim$300 supernovae from the presently running Nearby
Supernovae Factory \cite{snf} in the bin with lowest redshift.

Complemented by the Planck mission, a satellite to observe the CMB
at $z=1089$ and thus  measure various cosmological parameters and
breaking degeneracies in the SNAP observations, respectable
constraints can be obtained on the equation of state of dark
energy and its evolution with time.

SNAP's wide-field camera is not limited to studying supernovae,
however. It will also  be able to make use of the new, rapidly
emerging observational method of weak gravitational lensing (SNAP
[WL]) in mapping out the time evolution of the scale factor of the
universe $a(t)$, including through the growth factor of large scale
structure. Lensing will provide an independent measurement of the
evolution history and through complementarity allow even tighter
constraints.

Determining the equation of state is the crucial observational
clue for the nature of  dark energy. Commonly one defines the
equation of state function $w(z)$ by \BE p=w(z)\rho. \EE In
general, it is a function that can vary with time or redshift
$z$. For pure vacuum energy, $w=-1$ independent of redshift.

To link observational data sets to existing physical models, it
is useful to use a fitting  function for $w(z)$ that contains
only a few fitting parameters. The fewer parameters, the better
their value can be constrained by a given observational data set,
but---depending on the data---the less the resulting function
might fit the actual data.

A widely applicable fit, combining the virtues of having only two
fit parameters ($w_0$ and $w_a$),  yet fitting many theoretically
conceivable scalar field potentials, especially in the slow roll
regime, has been in use in cosmology for some time now, ever since
its introduction in  \cite{linwa}: \BE \label{fit}
w(z)=w_0+w_a\frac{z}{1+z} \ . \EE In particular it fits well small
recent deviations from vacuum energy with $w=-1$. Most models of
dark energy, except for some with heavily oscillating behavior
like some PNGB models (cf.\ e.g.\ \cite{Weller:2001gf}) approach
the limit of being barely distinguishable from the cosmological
constant in a manner compatible with Linder's fit (\ref{fit}). For
example, the linear potential treated in \cite{Kallosh:2003bq}
shows a deviation from $w=-1$ to slightly higher values of $w$
only for $z\la1$. Of course, eventually in the future, once people
have obtained the actual measurement data set, they will be able
to compare it to the cosmological constant model directly, not
having to revert to a fit first.

We take the SNAP baseline mission, as described in \cite {klmm},
including statistical and  systematic errors amounting to a
distance uncertainty of 1\% at the depth $z=1.7$ of the survey.
We marginalize over the absolute magnitude parameter $\mathcal
{M}$, which includes the Hubble constant $H_0$, and over
the dark energy density
$\Omega_D$, where we assume the preferred value to be centered at
$\Omega_D=0.72$, as favored by current observations (e.g.\
\cite{Spergel:2003cb}). Complementary data
from Planck, or from SNAP[WL], make it unnecessary to impose a
prior on $\Omega_D$.
The degeneracies are broken well enough to determine $\Omega_D$
with a precision comparable to a prior of
$\sigma_{\Omega_D}=0.01$ \cite{fhlt}.

 The constraints from the data on the dark energy fit
parameters $w_0$ and $w_a$ are analyzed  within the Fisher matrix
method \cite{tegfish, tth}, providing probability ellipses at
selected levels of confidence. Throughout this paper, we have
chosen the 95\% (or 2$\sigma$) confidence level. A much more
detailed account on specifics of SNAP, its observational
properties, errors and statistics, as well as a practical guide
for reproduction of the Fisher matrix method used here, is given
in \cite{Kallosh:2003bq} and references therein.

\begin{figure}[h!]
\centering\leavevmode\epsfysize= 5.4 cm
\epsfbox{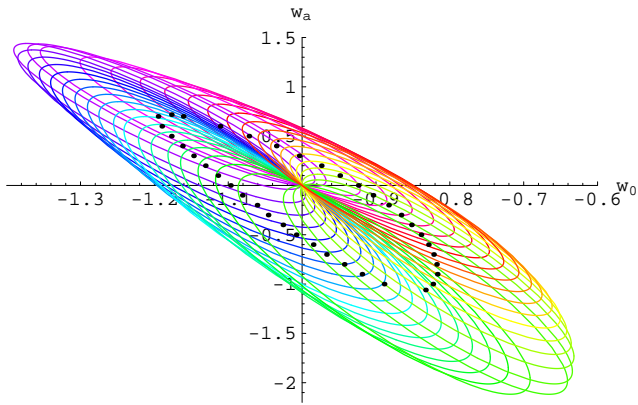}

\

\caption[fig1] {38 Fisher ellipses conspiring at 95\% confidence
to outline the region beyond which the cosmological constant is ruled
out as the main component of dark energy: if the most likely parameter value point
from the combined SNAP [SN] and Planck measurement will lie
outside the contour delineated by the centers (black dots) of the colored
ellipses, we will have to abandon the idea of dark energy being
due to pure vacuum energy with equation of state $p=-\rho$. This same black
contour is also depicted by the outer (blue) curve in Fig.~\ref{Contours}.}
\label{SNAP_SN-Planck}
\end{figure}

Fig.\ \ref{SNAP_SN-Planck} depicts the 95\% confidence level
contour as obtained from SNAP [SN]  and Planck, beyond which the
cosmological constant can be excluded from being responsible for
the dark energy density. If the SNAP supernovae measurements and
the CMB results from the Planck mission select a point in this
parameter space as being the most likely that lies outside of this
contour, then the dark energy density causing the present
acceleration of the expansion of the universe originates from
something different than vacuum energy.

Notice that all the ellipses in Fig.\ \ref{SNAP_SN-Planck} exactly
touch the cosmological constant  point ($w_0=-1$, $w_a=0$), the
criterion for obtaining the contour. It is clear from the
construction that the figure requires pointwise interpretation: as
the Fisher ellipse is supposed to be centered around the actual
(future) measurement point, not all the ellipses in the figure can
be valid at the same time. Only one will be eventually, after the
measurements will have been made, and in fact likely not even one
that is drawn. What the graph tells us is that any point outside
the contour, if the one most favored by the measurements, will
have an associated ellipse not encompassing the cosmological
constant point ($w_0=-1$, $w_a=0$), thus enabling us to exclude
the cosmological constant, at the 95\% confidence level, as a
possible cause of dark energy. Conversely, if the future
measurement point will happen to lie inside the contour, we will
be unable to rule out the cosmological constant, because the
ellipse will encompass the point (-1, 0).

To clarify our methodology, let us note that instead of drawing
many ellipses touching the point $w_0=-1$, $w_a=0$, and then
connecting their central points, one could draw, as usual, one
ellipse corresponding to the 2$\sigma$ deviation from this point.
This would give an ellipse somewhat similar to the contour
discussed above. However, this ellipse would be slightly
different, and it would have a different interpretation. It would
show us the points in the parameter space ($w_0$, $w_a$) excluded
at the 2$\sigma$ level by the observations favoring the simple
cosmological constant model. This is not what we study in our
paper.

Another comment should clarify the way we are using the
parametrization $w_0$, $w_a$, related to the fit (\ref{fit}). This
fit allows one to describe a broad class of the models of dark energy
(including the simplest cosmological constant), but there might
exist some exotic models for which this fit is inadequate. It is
important to understand that when we will have the real data from
SNAP and Planck, we will not need to use any fit anyway, as we may
directly compare the data with the predictions of the simplest
$\Lambda$CDM model. However, the use of the broadly applicable fit
(\ref{fit}) allows one to obtain a very good idea of the results
that can be obtained by various experiments.

In particular, SNAP will also carry out a wide field weak
gravitational lensing survey.  Measurements of the distortion of
distant source shapes by intervening gravitational potentials
probe the cosmology through both geometric effects on distances
and the growth of large scale structure.  Fig.~\ref{Contours} adds
the expected constraints from this weak lensing information (see
Appendix for more details) to the supernova and cosmic microwave
background data.

 \begin{figure}[h!]
\centering\leavevmode\epsfysize= 4.5 cm
\epsfbox{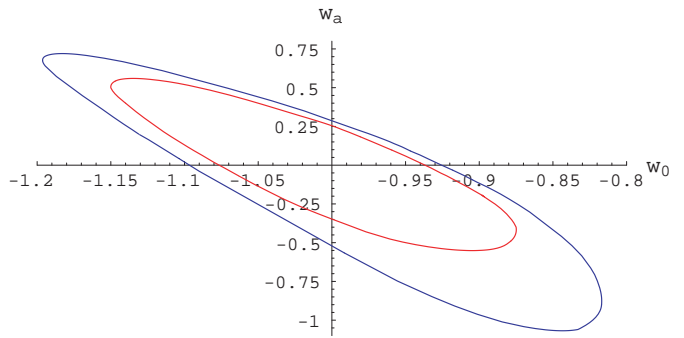}

\

\caption[fig2] {The contours outside of which the cosmological
constant can be excluded as the dominant contribution to dark energy.
The outer contour
(blue) was obtained by taking into account the observational systematic and
statistical uncertainties of SNAP [SN] and Planck. For the inner contour
(red), additionally SNAP [WL] was included.}
\label{Contours}
\end{figure}

Due to complementarity in the parameter dependencies, addition of
weak lensing data can significantly improve constraints and narrow
the region compatible with the cosmological constant.
Fig.~\ref{Contours} shows the 95\% confidence level contour
defined in Fig.~\ref{SNAP_SN-Planck} (outer, blue curve) by the
ensemble of possible measurement realizations and also the
corresponding (inner, red) curve when supplemented with weak
lensing information.  The phase space consistent with the
cosmological constant is reduced by a factor of two.  These data
should be available within the next 10 years.  Even tighter
constraints may be possible by including other future cosmological
observations, such as LSST's nearly full sky weak lensing survey a
few years later \cite{LSST-disc}.

A new, innovative method that has recently been explored in the
context of constraining $w(z)$ \cite{Garriga:2003nm}, is
cross-correlating CMB anisotropies with the matter power spectrum.
However, as pointed out in \cite{Garriga:2003nm}, the
corresponding error bars today are still far too big.  It is not
obvious whether cosmic variance does not spoil the results
obtained by this method irrecoverably \cite{Afshordi-privcom}.
This will require further investigation.

\

\section{Conclusions}

In a time when theoretical attempts at explaining the value of
$\Lambda$ have so far been of limited success, apart from
anthropic arguments
\cite{Linde1986,Weinberg:2000yb,Garriga:1999bf,Linde:2002gj}, it
is more important than ever to take the observational clues we
will have at our disposal to discriminate whether vacuum energy
poses the dominant contribution to the dark energy, or whether
different physics resides at its core.

It is an overzealous ambition to expect from present and future
observational  data that they will direct us unambiguously towards
a unique dark energy model. Yet, we can test each of the models
and rule out many of them. As we have shown in this paper, within
a decade, we shall gain the possibility to test and maybe even
rule out the most traditional of all dark energy models, the
cosmological constant model.

This model is by far the simplest of all models of dark energy.
Moreover, even though in the future we may learn how various dark
energy models could be related to string theory and M-theory, at
present the cosmological constant model remains the only one for
which this possibility was actually demonstrated
\cite{Kachru:2003aw}. Therefore measurements ruling out the
cosmological constant would have profound implications for
particle physics.

Meanwhile, measurements consistent with a cosmological constant
model will not resolve the mystery of the underlying
physics---whether it is a pure vacuum energy or a more complex
extension of physics merely possessing parameter values close to
the cosmological constant.  But at least in some of these cases
(as shown in \cite{Kallosh:2003bq}) we will have a grace period of
tens of billions of years to resolve the issue.

\medskip

As a program for the future, for the planning of observations
exploring the nature of dark energy, we would like to stress that
it is most important to realize the benefit obtained from the
interplay of \emph{various different} observation missions and
techniques. Degeneracies inherent in individual observation
methods are broken efficiently by considering several different
ones. If one takes, for instance, SNAP supernovae alone (contours
not depicted in this paper, but see \cite{Kallosh:2003bq} for
ellipses for that data set alone), the constraints are
informative, yet they gain a tremendous improvement from including
the Planck data coming from the measurements of the CMB, which
will be available around the same time as SNAP. Quite a further
impressive improvement is achieved if weak lensing from
SNAP---obtained by the same mission, but with a different
technique---is added. And another, equally impressive improvement,
although some years later than the above, is expected to come from
including LSST, a ground-based, nearly full sky weak lensing survey.

As we see, a lot of work to be done lies ahead, and only the joint
effort of a diversity of space- and ground-based observations,
combined with an innovative use and analysis of the data gathered
by these instruments, will provide us with the best possible
information on the nature of dark energy, fundamental cosmological
physics, and thereby with knowledge about the future
evolution and ultimate fate of our universe.

\bigskip

\bigskip

It is a pleasure to thank Niayesh Afshordi, Gary Bernstein, JoAnne Hewett, and
Anthony Tyson for useful discussions. The work by J.K.\ was
supported by the Stanford Graduate Fellowship and the Sunburst
Fund of the Swiss Federal Institutes of Technology (ETH Zurich and
EPFL). The work by A.L.\ was supported by NSF grant
PHY-0244728. The work by M.S.\ was supported by DOE grant
DE-AC03-76SF00515. The work by E.L.\ was supported in part by the
Director, Office of Science, DOE under DE-AC03-76SF00098 at LBL.

\

\section{Appendix}

\subsection{Weak Lensing with SNAP}

Gravitational lensing involves the deflection of light
rays from distant sources by the intervening gravitational
potentials of, e.g., large scale structures of matter.
One can relate the power spectrum of matter density fluctuations
to the distribution of image distortions (shear) and size
magnifications (convergence), weighted by distance dependent
strength or focusing factors.  Weak lensing refers to the regime
where these effects are small and observed statistically rather
than through more obvious multiple imaging, for example.  Thus
measurements of weak lensing probe the dark energy through the
expansion history, both its effect on distances and the suppression
of growth of matter structure as dark energy dominates and the
expansion accelerates.

Numerous methods involving weak lensing are actively being
investigated to develop an optimal probe of the nature of
dark energy.  The situation is still uncertain: only recently have
computations included a time
variation such as $w_a$ and systematic effects are not well
identified.  Here we
take a greatly simplified version of the most promising method
for reducing the dominant systematic error---cross-correlation
cosmography \cite{Bernstein-Jain, Jain-Taylor}.

This involves studying background sources at different redshifts
relative to the same foreground mass screen, allowing cancellation
of many issues such as spurious distortion due to the instrumental
point spread function and ignorance of the exact lensing mass
distribution.  This method comes close to providing a pure
geometric test of the ratio between comoving distances to the
various source redshifts.  While the true situation is more complex,
for the calculation in this paper we take SNAP wide field data
to provide determination of the ratio
\begin{equation}
R \equiv \frac{r_{s1}-r_l}{r_{s2}-r_l},
\end{equation}
where $r_l$ is the comoving distance to the lens and $r_{si}$ is
the comoving distance to the $i$th source plane,
to 0.2\% at three lens
redshifts $z_l=0.3$, 0.6, 0.9.  Furthermore we fix the sources
to lie in planes at $z_{s2}=2z_{s1}=4z_l$.  This is clearly a toy
model but it succeeds in reproducing the more sophisticated
parameter estimation contours in Fig.~2 of \cite{Bernstein-Jain}.
Thus this approach provides a readily calculable first step toward
including forthcoming weak lensing data.

\end{document}